\documentclass[a4paper,12pt]{article}

\begin{document}

\centerline {\bf ANOTHER PROOF OF GELL-MANN AND LOW'S THEOREM}

\vskip 0.5truecm

\centerline {Luca Guido Molinari}
\centerline {Dipartimento di Fisica, Universit\`a degli Studi di Milano,}
\centerline {and INFN, Sezione di Milano, Via Celoria 16, 20133 Milano, Italy}
\centerline {luca.molinari@mi.infn.it}

\vskip 1truecm

\centerline {Abstract}
The theorem by Gell-Mann and Low is a cornerstone in QFT and zero-temperature 
many-body theory. The standard proof is based on Dyson's time-ordered 
expansion of the propagator; a proof based on exact identities for the 
time-propagator is here given.
\vskip 0.5truecm


\vskip 1truecm

In the appendix of their paper ``Bound States in Quantum Field Theory'', 
Murray Gell-Mann and Francis Low proved a fundamental theorem\cite{Gellmann51}
that bridges the ground states $|\Psi_0\rangle $ and 
$|\Psi\rangle $ of Hamiltonians $H_0$ and $H=H_0+gV$ by means of 
time-propagators, and makes the transition of time-ordered
correlators from the Heisenberg to the Interaction picture possible:
\begin{eqnarray}
\langle \Psi |T \psi(1)\ldots \psi^\dagger (n)|\Psi\rangle =
\frac{\langle \Psi_0|T S \psi(1)\ldots \psi^\dagger (n)|\Psi_0\rangle }
{\langle \Psi_0|S|\Psi_0\rangle } \label{S}
\end{eqnarray}
The single operator $S=U_I(\infty,-\infty)$ contains all the effects of 
the interaction. The theorem borrows ideas from the scattering and the 
adiabatic theories and
introduces the concept of {\sl adiabatic switching} of the interaction, 
through the time-dependent operator
\begin{eqnarray}
H_\epsilon (t) = H_0 +  e^{-\epsilon |t|} gV. \label{HAM}
\end{eqnarray}
that interpolates between the operators of interest, $H$ at $t=0$ and
$H_0$ at $|t|\to\infty $. The adiabatic limit is obtained for 
$\epsilon\to 0^+$. 
With the operator $H_0$ singled out, the theorem requires the 
time-propagator in the interaction picture, 
\begin{eqnarray}
 U_{\epsilon I}(t,s) =e^{\frac{i}{\hbar} tH_0}U_\epsilon
(t,s)e^{-\frac{i}{\hbar} sH_0} \label{UI}
\end{eqnarray}
where $U_\epsilon (t,s)$ is the full propagator\cite{Reed}. The statement of 
Gell-Mann and Low's theorem is: 
\vskip 0.5truecm

\underbar {Theorem}. Let $|\Psi_0\rangle $ be an eigenstate 
of $H_0$ with eigenvalue $E_0$, and consider the vectors 
\begin{eqnarray}
|\Psi^{(\pm)}_\epsilon \rangle  = \frac{U_{\epsilon I}(0,\pm\infty )
|\Psi_0\rangle }{\langle \Psi_0 | U_{\epsilon I}(0,\pm\infty )|\Psi_0\rangle } 
\label{TH1}  
\end{eqnarray}
If the limit vectors $|\Psi^{(\pm )}\rangle $ for $\epsilon \to 0^+$ exist, 
then they are eigenstates of $H$ with same eigenvalue $E$. 
\vskip 0.5truecm

Despite the validity of the theorem beyond pertubation theory, in the original 
paper\cite{Gellmann51} and in textbooks\cite{Nozieres,Fetter,Gross} the proof 
makes use of Dyson's expansion of the interaction propagator, and is rather 
cumbersome. Mathematical proofs are based on the adiabatic 
theorem\cite{Nenciu}.
In this letter a simple alternative proof is given, based on exact
identities for the propagator that are stated in the following Lemma.

\vskip 0.5truecm 

{\underbar {Lemma}}. If $U_\epsilon (t,s)$ is the time-propagator 
for $H_\epsilon (t)$ then, for all positive $\epsilon $, the following 
relations hold:
\begin{eqnarray}
&&i\hbar \epsilon g\frac{\partial}{\partial g} U_\epsilon (t,s) =\nonumber\\ 
&&H_\epsilon (t) U_\epsilon (t,s) - U_\epsilon (t,s)H_\epsilon (s), \quad
{\rm if}\quad 0\ge t\ge s,  \label{L1}\\
&&-H_\epsilon (t) U_\epsilon (t,s) + U_\epsilon (t,s)H_\epsilon (s), \quad
{\rm if}\quad t\ge s\ge 0.  \label{L3}
\end{eqnarray}
{Proof}: The trick is to make the $g$-dependence of the 
propagator explicit into the time-dependence of some related propagator. 
Schr\"odinger's equation 
\begin{eqnarray}
i\hbar \partial_t U_\epsilon(t,s)=H_\epsilon(t) U_\epsilon (t,s),
\quad U_\epsilon(s,s)=1 \label{Schr}
\end{eqnarray}
corresponds to the integral one, where we put $g=e^{\epsilon\theta}$:
\begin{eqnarray}
U_\epsilon (t,s) = I + \frac{1}{i\hbar } \int_s^t dt' (H_0+ e^{\epsilon 
(\theta -|t'|)}V ) U_\epsilon (t', s) \label{Int1}
\end{eqnarray}
Consider the $g$-independent operators
$ H^{(\pm)}(t)= H_0+ e^{\pm \epsilon t}V$, with corresponding
propagators $U^{(\pm )}(t,s)$. 
For $0\ge t\ge s$, a time-translation in eq.(\ref{Int1}) gives
\begin{eqnarray}
U_\epsilon (t,s) = I + \frac{1}{i\hbar } \int_{s+\theta} ^{t+\theta } dt' 
H^{(+)} (t' ) U_\epsilon (t'-\theta , s) \label{int}
\end{eqnarray}
Comparison with the equation for $U^{(+)}(t+\theta, s+\theta)$:
\begin{eqnarray}
U^{(+)}(t+\theta, s+\theta)=I+\frac{1}{i\hbar } \int_{s+\theta} ^{t+\theta } 
dt' H^{(+)} (t' ) U^{(+)} (t', s+\theta) \nonumber
\end{eqnarray}
and unicity of the solution imply the identification
\begin{eqnarray}
U_\epsilon (t,s) = U^{(+)}(t+\theta, s+\theta )
\end{eqnarray}
Since $\theta $ enters in the operator $U^{(+)}(t+\theta,s+\theta )$ 
only in its temporal variables, we obtain
\begin{eqnarray}
\partial_\theta U_\epsilon (t,s) = \partial_t U_\epsilon (t,s) + 
\partial_s U_\epsilon (t,s) 
\end{eqnarray}
By using eq.(\ref{Schr}) and its adjoint, the first identity is proven.

If $t\ge s\ge 0$, the same procedure gives $U_\epsilon (t,s)= U^{(-)} 
(t-\theta, s-\theta )$ and therefore
$\partial_\theta U_\epsilon (t,s) = -\partial_t U_\epsilon (t,s)- 
\partial_s U_\epsilon (t,s)$, which leads to the identity (\ref{L3}). An 
identity for $t\ge 0\ge s$ can be obtained by writing $U_\epsilon (t,s)=
U_\epsilon (t,0)U_\epsilon (0,s)$.
\vskip 0.5truecm
 
In the interaction picture, eq.(\ref{UI}), the identities
transform straightforwardly into the following ones:
\begin{eqnarray}
&&i\hbar  \epsilon g\frac{\partial}{\partial g} U_{\epsilon I}(t,s) =
\label{UI13}\\
&&H_{\epsilon I}(t) U_{\epsilon I}(t,s)- U_{\epsilon I}(t,s)
H_{\epsilon I}(s), \quad  {\rm if}\quad 0\ge t\ge s, \nonumber \\
&&-H_{\epsilon I}(t)U_{\epsilon I}(t,s)+ U_{\epsilon I}(t,s)
H_{\epsilon I}(s),\quad {\rm if}\quad t\ge s \ge 0,\nonumber 
\end{eqnarray}
where $H_{\epsilon I}(t)=e^{\frac{i}{\hbar} tH_0} H_\epsilon (t)
 e^{-\frac{i}{\hbar} tH_0}$.
By applying eqs.(\ref{UI13}) to an eigenstate $|\Psi_0\rangle $ 
of $H_0$, we obtain 
\begin{eqnarray}
(H-E_0\pm  i\hbar \epsilon g \frac{\partial}{\partial g}) U_{\epsilon I}
(0,\pm\infty)|\Psi_0\rangle = 0 \label{L21}
\end{eqnarray}
This same equation is proven in the literature by direct use 
of Dyson's expansion. From now on, the proof of the theorem proceeds in
the standard path, and is here sketched for completeness. 

Existence of a non-trivial adiabatic limit of eq.(\ref{L21}) requires 
the vector $U_{\epsilon I}(0,\pm \infty)|\Psi_0\rangle $ 
to develop a phase proportional to $1/\epsilon $; this has been checked
in diagrammatic expansion\cite{Hubbard,Kazuo91}. 
The singular phase is precisely 
removed by the denominator in the definition of the vectors 
$|\Psi^{(\pm)}_\epsilon \rangle $, before the limit is taken.

1)  For finite $\epsilon $, the two identities eq.(\ref{L21}) are 
projected on the vector $|\Psi_0\rangle $, and yield a formula for
the {\sl energy shift}
\begin{eqnarray} 
\mp i\hbar \epsilon g \frac{\partial}{\partial g} \log 
\langle \Psi_0| U_{\epsilon I}(0,\pm\infty)|\Psi_0\rangle = 
E^{(\pm)}_\epsilon  -E_0 \label{L22}
\end{eqnarray}
where $E_\epsilon^{(\pm )}= \langle \Psi_0|H|\Psi_\epsilon^{(\pm )}\rangle $.

2) By eliminating $E_0$ in (\ref{L21}) with the aid of (\ref{L22}), 
with simple steps one obtains
\begin{eqnarray}
(H-E_\epsilon^{(\pm)}\pm i\hbar\epsilon g\frac{\partial}{\partial g}) 
|\Psi^{(\pm)}_\epsilon \rangle = 0\label{Lem2}
\end{eqnarray}
Then, for $\epsilon \to 0^+$, the limit vectors $|\Psi^{(\pm)}\rangle $
obtained by pulling onward or backward in time the same asymptotic
eigenstate $|\Psi_0\rangle $, are eigenvectors of $H=H_0+gV$ with 
eigenvalues $E^{(\pm)}$.

3) The time-reversal operator has the action $T^\dagger U_\epsilon (t,s) T= 
U_\epsilon (-t,-s) $. If $H_0$ commutes with $T$ the relation extends to
the interaction propagator and $T^\dagger U_{\epsilon I}(0,\infty) T = 
U_{\epsilon I} (0,-\infty)$. If $|\Psi_0\rangle $ is also an eigenstates of 
$T$, it follows that $T^\dagger |\Psi^{(+)}_\epsilon \rangle $ is parallel to 
$|\Psi^{(-)}_\epsilon \rangle $ and $E^{(+)}= E^{(-)}$.
The proportionality factor equals one, since 
$\langle E_0|\Psi^{(+)}\rangle =\langle E_0|\Psi^{(-)}\rangle $.

The formula for the energy shift, eq.(\ref{L22}), can be recast in a form 
involving the $S$-operator (Sucher's formula)\cite{Sucher}. From 
eqs.(\ref{UI13}) the following relation follows:
\begin{eqnarray}
-i\hbar \epsilon g \partial_g\, S = H_0S + SH_0 - 2U_{\epsilon I}(\infty, 0)
H U_{\epsilon I}(0,-\infty)\nonumber
\end{eqnarray}
The expectation value on the eigenstate $|\Psi_0\rangle $ and use of the 
theorem give Sucher's formula:
\begin{eqnarray}
E-E_0 = \lim_{\epsilon\to 0} \frac{i\hbar\epsilon}{2} g\frac{d}{dg}\log
\langle \Psi_0|U_{\epsilon I}(\infty, -\infty)|\Psi_0\rangle 
\end{eqnarray}

\end{document}